# High Responsivity and Quantum Efficiency of Graphene / Silicon Photodiodes Achieved by Interdigitating Schottky and Gated Regions

Sarah Riazimehr[a], Satender Kataria[a*], Jose-Maria González-Medina[b], Mehrdad Shaygan[c], Stephan Suckow[c], Francisco G. Ruiz[b], Olof Engström[c], Andres Godoy[b], Max Christian Lemme[a,c*]

[a]RWTH Aachen University, Chair of Electronic Devices, Faculty of Electrical Engineering and Information Technology, Otto-Blumenthal-Str. 2, 52074 Aachen, Germany

[b]Dpto. de Electrónica y Tecnología de Computadores, Facultad de Ciencias, Universidad de Granada, Av. Fuentenueva S/N, 18071 Granada, Spain

[c]AMO GmbH, Advanced Microelectronic Center Aachen (AMICA), Otto-Blumenthal-Str. 25, 52074 Aachen, Germany

*email: satender.kataria@eld.rwth-aachen.de, max.lemme@eld.rwth-aachen.de

## Abstract

Graphene / silicon (G/Si) heterostructures have been studied extensively in the past years for applications such as photodiodes, photodetectors and solar cells, with a growing focus on efficiency and performance. Here, a specific contact pattern scheme with interdigitated Schottky and graphene/insulator/silicon (GIS) structures is explored to experimentally demonstrate highly sensitive G/Si photodiodes. With the proposed design, an external quantum efficiency (EQE) of > 80% is achieved for wavelengths ranging from 380 to 930 nm. A maximum EQE of 98% is observed at 850 nm, where the responsivity peaks to 635 mA/W, surpassing conventional Si p-n photodiodes. This efficiency is attributed to the highly effective collection of charge carriers photogenerated in Si under the GIS parts of the diodes. The experimental data is supported by numerical simulations of the diodes. Based on these results, a definition for the 'true' active area in G/Si photodiodes is proposed, which may serve towards standardization of G/Si based optoelectronic devices.

## Introduction:

Graphene optoelectronic devices typically show limited photoresponsivity due to weak optical absorption per layer (2.3%)[1]. Nevertheless, the material has great potential as a transparent and anti-reflective junction when combined with two-dimensional (2D) and three-dimensional (3D) semiconductors for a number of optoelectronic device applications [2–4]. Graphene/silicon (G/Si) is the most widely studied fundamental heterostructure, as it extensively utilizes the existing Si technology [2,5,6]. G/Si heterostructures are a basic component of many potential applications, such as rectifiers[7], chemical and biological sensors[8–10], solar cells[11–14] and - in particular - photodetectors[3,4,15–28]. The latter include G/Si Schottky photodetectors for near infrared (NIR)[25] and ultraviolet (UV)[29] light detection. Different strategies have been explored to enhance the performance of G/Si photodiodes. Chemical doping of graphene has been demonstrated by Miao *et al.* to shift the Fermi level in graphene down with respect to the Dirac point, thereby increasing the built-in potential in G/Si solar cells and resulting in an EQE of 65%[14]. However, chemical doping of graphene is unstable and unreliable when exposed to air and humidity[30]. Another approach is to utilize interfacial oxide layers on photodiodes. Selvi *et al.* demonstrated that such interfacial oxides increase the Schottky barrier height (SBH) and lead to an enhancement in photovoltage responsivity, particularly for low light intensities[31]. Complex nanotip patterning of substrates was employed to improve the photoresponse of G/Si Schottky diodes[28]. Here, the nano-tip surface enhances light collection due to multiple reflections, and the tip-enhanced electrical field enables photo-carrier separation with internal gain due to impact ionization, resulting in responsivity of 3 A/W under white light illumination[28]. Recently, we have pointed out the significant role of insulating regions in parallel to the Schottky regions in achieving high photocurrents through scanning photo current measurements[3]. We have further provided a model for the physical mechanisms, i.e. the formation of an inversion layer in Si[3]. Similar conclusions have been reported in[21,23], where capacitance-voltage and current-voltage (I-V) measurements were carried out at different temperatures for validation. This fundamental observation of origins of photocurrent generation is a key to further improve the performance of G/Si photodiodes.

In this work, we demonstrate that the graphene on interdigitated patterned $SiO_2$/Si substrates harvests the carriers photogenerated in Si much more efficiently than the conventional G/Si diode structures, where only small areas of $SiO_2$ are used just to isolate graphene from Si in the non-Schottky regions. The investigated structures are compatible with the existing Si technology and require no specific process steps for fabricating G/Si diodes beyond graphene transfer. Our devices show an absolute spectral responsivity of 635 mA/W (calculated by considering the entire charge generation area, which includes oxide regions). This value is about 30% higher than that observed in commercial Si photodetectors and about 50% higher than recently reported for Si p-i-n photodiodes with complex photon-trapping microstructures at a wavelength of 850 nm. [32]

## Results and discussions

Motivated by our previous findings,[3] diodes with interdigitated G/Si (Schottky) and graphene/insulator/silicon (GIS) regions were designed, where the insulated regions made up a significant portion of the active device area. Two different types of diodes were fabricated per chip. In diode D1, graphene is in contact with interdigitated $SiO_2$/Si structured substrates (Fig. 1a), while in diode D2, the control device, graphene is mostly in contact with the Si substrate similar to conventional diode layouts and covers just a small $SiO_2$ area (Fig. 1b). A number of D1-type diodes were designed with varying widths of the GIS regions (supplementary Fig. S1). The devices were fabricated by transferring large-area CVD graphene onto pre-patterned substrates of lightly doped n-Si with a $t_{ox}$ = 20 nm $SiO_2$ layer (details in Methods section), and can, in principle, be carried out at the wafer scale[33]. The final chips with different devices were inserted and wire-bonded into a chip carrier for optoelectronic characterization (Fig 1c). Fig. 1d and 1e show color enhanced scanning electron micrographs of diodes D1 and D2, respectively. The two distinct regions in each device, i.e. the G/Si Schottky junctions and the GIS structures, behave differently under illumination. However, charge carriers can only be extracted through the Schottky junction areas, which act as a collector. This region determines basic diode parameters such as ideality factor and SBH. The graphene quality was assessed through scanning Raman spectroscopy using a wavelength of 532 nm. Fig. 1f shows an optical micrograph of the region where Raman spectra of graphene on Si and on $SiO_2$ was acquired. A 2D/G intensity ratio >1 indicated the monolayer nature of CVD graphene transferred on the devices. We did not observe changes in Raman spectrum of graphene located on Si or $SiO_2$ beyond a generally lower intensity of the graphene Raman peaks on Si, as expected (Fig. 1g, 2D band intensity).

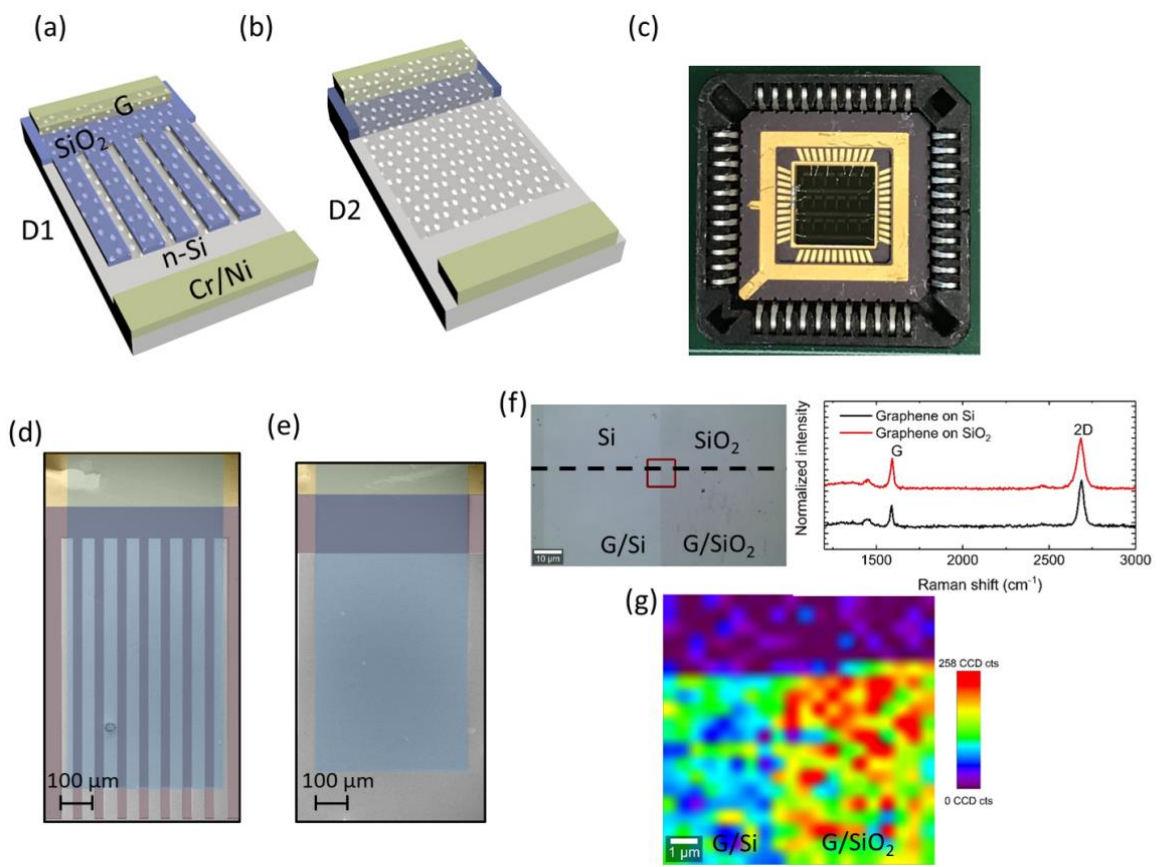

*Fig. 1: Schematic of G/Si photodiodes of type (a) D1 and (b) D2 (control device). (c) Photograph of a wirebonded diode chip in a chip package. Scanning electron micrograph of two diodes of type (d) D1 and (e) D2. The SEM image is color enhanced to show the position of the graphene film (blue), $SiO_2$ (purple), and the metal electrodes (yellow). (f) Optical*

*micrograph of the region selected for acquiring a Raman area map (red square). Graphene covers the area below the dashed line. Single Raman spectra of graphene on Si (lower curve) and on SiO₂/Si region (upper curve) acquired for λ = 532 nm. (g) Raman intensity map of the 2D band in the area marked in (f).*

Current-voltage (I-V) characteristics of the diodes under dark (solid lines) and illuminated (dashed lines) conditions are shown in Fig. 2a in semi-logarithmic scale. Both the diodes D1 and D2 clearly exhibit rectifying behavior in the dark with rectification ratios, defined as forward current @ +2 V) / reverse current @ -2 V, of up to $5\times10^4$ and $4\times10^4$, respectively. The general photoresponsivity of the diodes was measured under LED white-light illumination with an intensity of 30 µW/cm² (see Methods). The forward I-V characteristics of the diodes were fitted using standard expressions[34,35], which yielded ideality factors (η) of 2.08 and 2.3 and Schottky barrier heights (SBH) of 0.76 and 0.79 eV for diodes D1 and D2, respectively. Only the Schottky (collection) area has been employed for the estimation of these parameters. The ideality factors (η) reported for G/n-Si Schottky junctions varies in the range of ~1.1 to 7.69, depending on the quality of the interface between the graphene and the semiconductor[2,14,36]. A significant variation in reported values of η can be attributed to inhomogeneities and impurities at the G/Si junction[37] originating from graphene-transfer related issues[38] or thin inevitable interfacial SiO₂ layers[2,31,39]. A small photovoltaic effect can be observed in Fig. 2a for both devices. A slightly higher open circuit voltage in D2 (112 mV) compared to D1 (73 mV) can be attributed to a higher built-in potential in D2, as the SBH is greater in this device. It should be mentioned here that the SBH values in G/Si diodes strongly depend on the quality of the transferred graphene and on the interface properties between graphene and Si. Therefore, SBH values can significantly vary from device to device. The evolution of photocurrent, defined as $I_{ph} = I_{light} - I_{dark}$, versus reverse voltage (V_R) for both diodes is shown in figure 2b. Diode D1 exhibits higher photocurrent in most of the applied voltage range than the control device (D2), except below -0.6 V where the two curves cross and D2 yields higher photocurrent than D1. Similar bias voltage dependencies were observed in spectral response (SR) measurements of the devices (see below). This can be explained through the formation of an inversion layer under the GIS part of the devices, as shown for other oxide thicknesses in [3]. The formation of such an inversion layer decreases surface recombination at the $SiO_2$/Si interface[40,41]. It further forms a quasi PN junction with the regions covered by Schottky contacts, which facilitates the transit of the photogenerated carriers to the Schottky junctions[3,21,23]. The so formed inversion layer in the GIS region results in a higher photocurrent in diode D1 than in a purely Schottky junction of the same area (D2).

Large-area scanning photocurrent (SPC) measurements provide insight into the spatial distribution of photocurrents. We recorded SPC maps of the diodes using a $\lambda = 532$ nm laser with a spot size of 2.4 µm. Figures 2c and 2d show optical micrographs of both devices, where the red rectangle indicates the scanned area and the black dashed line represents the graphene area. The maps were obtained using a 10 × objective over an area of 1.3 mm × 0.6 mm with a scan speed of 0.4 s/line and integration time of 1 ms. Photocurrent maps of the devices obtained at -2 V with a laser power of 5 µW are shown in Fig. 2e and 2f. For both devices, higher photocurrents were recorded in the GIS region compared to the G/Si region. This observation is consistent with our previous findings[3], and attests the significance of the inversion layer in the GIS regions towards the total photocurrent. The devices even showed a significant photocurrent at low reverse biases, pointing towards their potential applicability as low power and highly sensitive photodiodes (supplementary Fig. S2). Moreover, it should be highlighted that the diodes did not show a significant degradation in their electrical performance, even without encapsulation, over a time span of nearly six months (supplementary Fig S3). The samples were kept in a simple vacuum desiccator during this time.

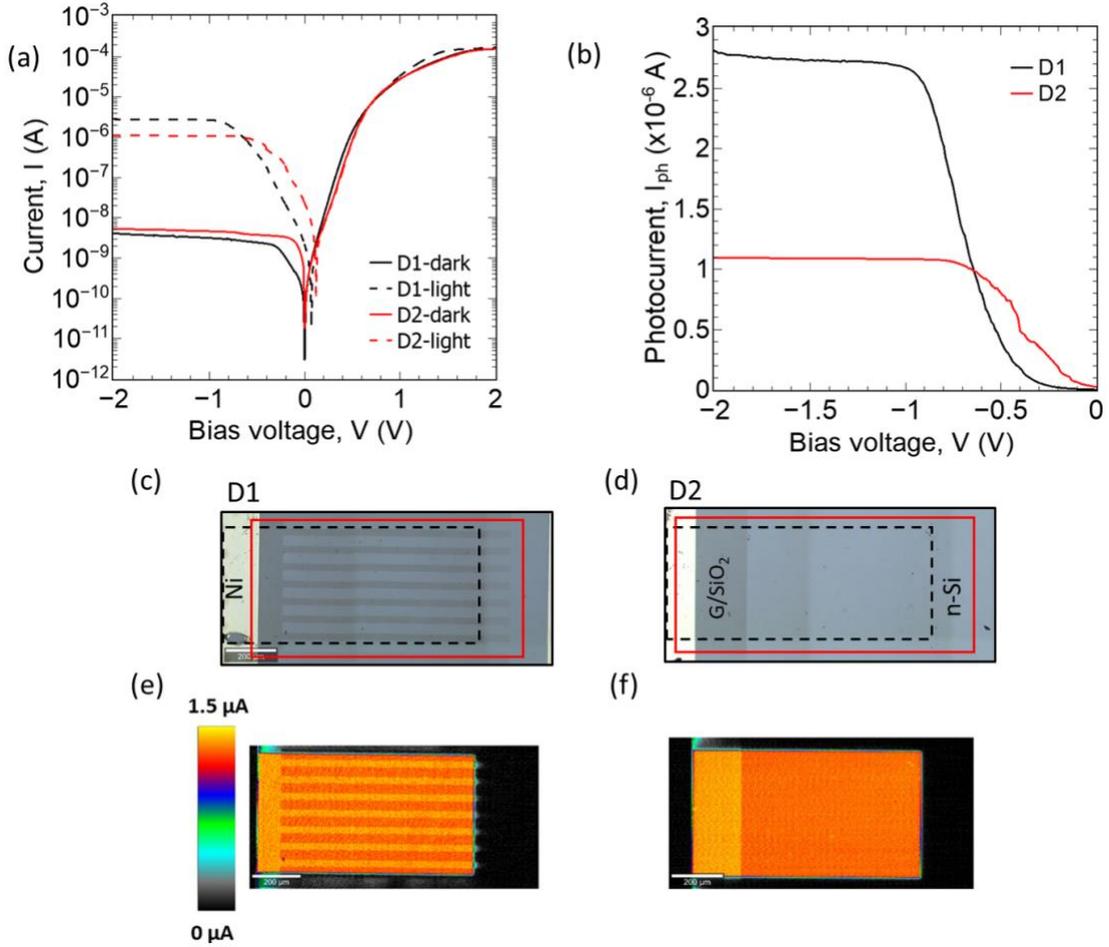

*Fig. 2: (a) Diode current in logarithmic scale as a function of bias voltage of G/Si photodiodes (D1 and D2) in the dark and under illumination. The effect of photocurrent generation can be observed for negative bias voltage, i.e. in reverse bias mode. (b) Photocurrent of G/Si photodiodes (D1 and D2). Optical micrograph of diodes (c) D1 and (d) D2. The area inside the red rectangle in (a) and (b) was scanned for photocurrent measurements. The graphene region is represented by the black dashed rectangle. Photocurrent maps of the scanned area for (e) D1 and (f) D2 at a laser wavelength of 532 nm, power of 5 µW and a reverse bias of -2 V. The scale bar in (c)-(f) corresponds to 200 µm.*

A figure of merit for photodetection, i.e. how a photodetector absorbs and responds to the incoming light, is responsivity. It is a measure of output current per unit of incident optical power. In conventional Schottky and PN photodiodes, the Schottky or PN region is considered as the active area for calculating the absolute SR or photoresponsivity. Here, "active area" is equivalent to the device's region, where charge carriers are generated by the absorbed incident photons. Conventional Schottky diodes consist of thin metal films in the active area and thick metal contacts, which are opaque to light. The latter regions are legibly not considered as photoactive areas. Graphene, in contrast, has broadband transparency of around 97%, which allows light to penetrate into the Si substrate in both the Schottky and the GIS areas. This enables illumination of silicon trough the G/SiO$_2$ stack and generation of charge carriers therein. This is also confirmed in SPC measurements (Fig. 2e and 2f) where significant photocurrent was detected underneath GIS regions because of light absorption in Si. Thus, both the GIS and the Schottky area have to be defined as active regions, which we propose to call "generation area" $A_{gen}$, rather than active area, with

$$A_{gen} = A_{GIS} + A_{GS}, \qquad (1)$$

where $A_{GIS}$ is the GIS area and $A_{GS}$ is the G/Si Schottky junction area, which is marked with a dashed black line in Fig. 3a. This consideration is of significant importance while measuring SR of the diodes where area consideration is warranted. The absolute SR was then measured taking into account $A_{gen}$, marked with a dashed orange line in Fig. 3a, through a comparative method to the known SR of a reference detector using a LabVIEW controlled setup. Details of the procedures are explained in the supporting information.

Figure 3b shows the SR of the diodes measured over a spectrum from λ = 360 to 1200 nm and for two different applied reverse bias voltages of -2 V (solid lines) and -0.5 V (dashed lines). For a comparison, the responsivity of a commercial IR-enhanced Si PIN photodetector (S1337-33BQ-Blue curve) at applied reverse bias of -2 V (at its normal operating point) is also shown in Fig. 3b. The responsivity values are reported by considering the proposed area given by Eq. 1. The highest absolute responsivity was obtained for the interdigitated device D1 with a maximum of 635 mA/W at λ = 850 nm, which is about 89% higher than that of D2 (maximum of 330 mA/W). It is further surpassing the SR of the commercial detector by about 28% (maximum of 490 mA/W) measured at -2 V under the same conditions. The enhanced SR of diodes with interdigitated structures was found to be consistent for different devices with varying GIS region widths (supplementary Fig. S1). In all cases, the maximum response is well within the Si absorption spectrum. Both devices D1 and D2 showed an increase of the absolute spectral responsivity with the applied reverse bias, but D1 showed a lower responsivity than D2 at an applied reverse bias of -0.5 V (Fig. 3b, black and red dotted lines). D1 then outperformed D2 for a larger bias of -2 V (Fig. 3b, black and red solid lines). This is in agreement with the measured $I_{ph}$ in Fig. 2b. To demonstrate how much responsivity can be overestimated, the SR was recalculated for both devices by only considering the Schottky area ($A_{GS}$) as an active area, which is obviously smaller than $A_{gen}$. The dashed lines in Fig. 3c represent overestimated responsivity, while the solid lines were calculated by considering the entire $A_{gen}$, resulting in correct responsivity. This clearly shows that ignoring the GIS area in G/Si diodes can lead to significant overestimation of responsivity, especially when the GIS area covers large parts of the device.

External quantum efficiency (EQE) of the devices was calculated from the measured SR as: $EQE = \frac{R}{q}\frac{hc}{\lambda}$. Here, $q$, $h$, $\lambda$ and $c$ are electron charge, Planck's constant, wavelength and speed of light, respectively. Figure 3d shows the EQE of devices D1, D2 and the commercial Si photodetector over a broad spectrum at an applied reverse bias of -2 V. The EQE in D1 reached values above 80% in the wavelength range from 380 nm to 930 nm, covering the whole visible and near infrared region, with a maximum value of 98%. As Si is the main absorber in these devices, EQE decays quickly for photon energies below the bandgap of Si. This very high EQE for the G/Si photodiodes has been achieved merely by interdigitating the GIS regions with the Schottky junctions, i.e. without complex device designs. This compares extremely well with pure Si p-i-n diodes, where an EQE of about 60% was achieved recently for a limited wavelength range from 800 to 860 nm, where complex

micro- and nanostructured holes were used for efficient light trapping [32]. With this design, we obtained a comparable EQE to that reported in [42], in which the authors used nano-structured black Si surfaces in combination with conformal alumina coating to enhance the responsivity of Si-based photodiodes.

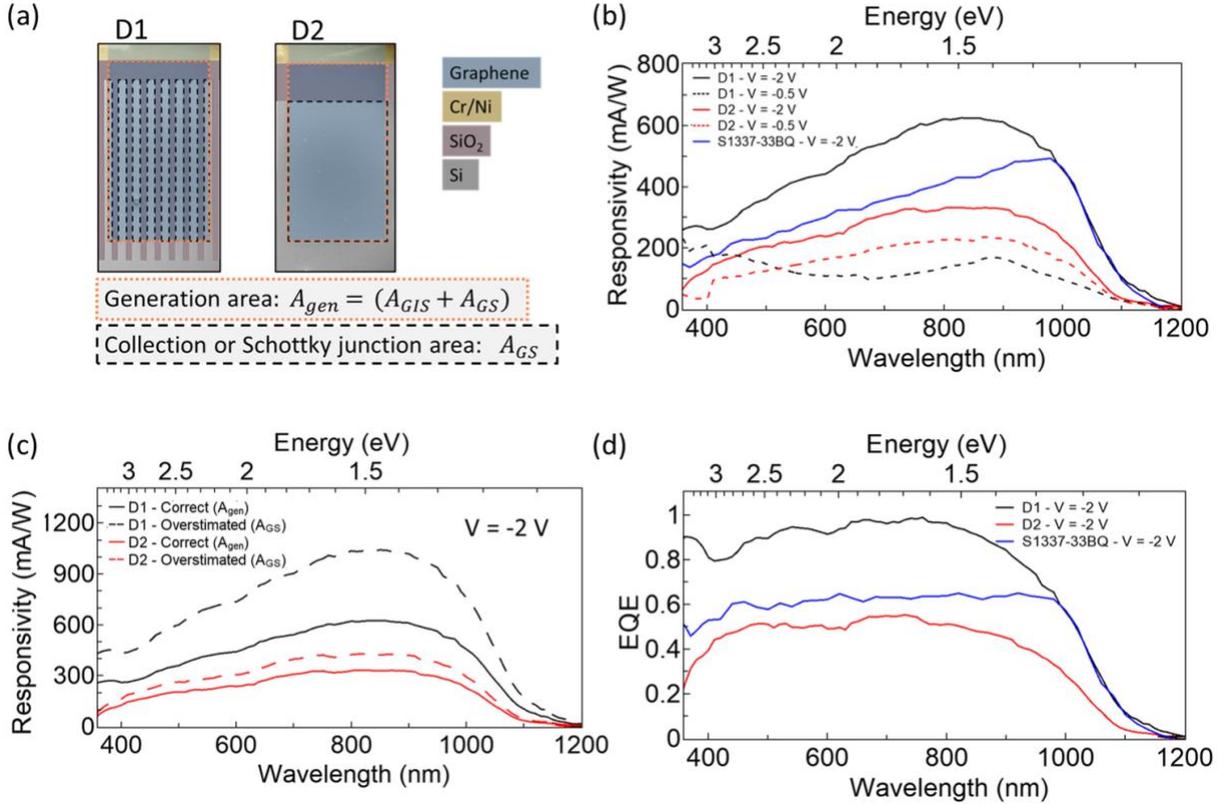

*Figure 3: (a) Color enhanced scanning electron micrographs indicating the photocurrent generation area ($A_{gen}$) and the collection area ($A_{GS}$) in devices D1 (left) and D2 (right). (b) Measured responsivity vs. wavelength (lower x-axis) and energy (upper x-axis) of G/Si photodiodes for wavelengths ranging from 360 nm (3.4 eV) to 1200 nm (1 eV) at reverse biases of -2 V (solid lines) and -0.5 V (dotted lines) in comparison with a commercially available Si ( S1337-33BQ) photodetector (blue line). (c) Overestimated responsivity (dashed lines) calculated by considering only $A_{GS}$ in comparison with correctly calculated responsivity by considering the entire $A_{gen}$(solid lines) of D1(black) and D2 (red) at reverse bias of -2 V. (d) External quantum efficiency (EQE) vs. wavelength and energy of G/Si photodiodes at reverse bias of -2 V in comparison with S1337-33BQ photodetector (blue line).*

**Physical model and simulation**

Numerical simulations were performed to understand the physics controlling the interdigitated structures, and in particular in the outstanding increase of $I_{ph}$, SR and hence the improvement in EQE. We have self-consistently solved the Poisson and continuity equations in a 2D section of the structures, shown between the vertical dashed lines in Fig. 4a, including the GIS and G/Si regions (see Methods). For the simulation, a uniform illumination over the device surface with a wavelength of $\lambda = 800$ nm and an intensity of $25\ \mu W/cm^2$ was considered. This wavelength value was selected as it falls in the spectral region where maximum SR and EQE were measured in the experiments (Fig. 3). An absorption coefficient of $\alpha = 10^3\ cm^{-1}$ was used in the Si-substrate for such a value of $\lambda$[33]. Fig. 4b shows the calculated electric field (E) in the selected region of the device. The highest values of E are located at the insulating $SiO_2$ and below the graphene. The electric field is directed towards the surface due to the negative applied bias to the graphene contact and the corresponding curvature of the bands along the z-axis. However, it should be highlighted that a relevant contribution of the lateral electric field ($E_x$) appears at the transition from the GIS to the G/Si region. This lateral component of E inside the Si substrate is represented in Fig. 4c at different depths from the Si surface, with z = 20 nm (the $SiO_2$ thickness) indicating the top Si position (G/Si and $SiO_2$/Si interfaces). Here, the lateral component of the electric field $E_x$ contributes only at the transition from GIS to the Schottky region.

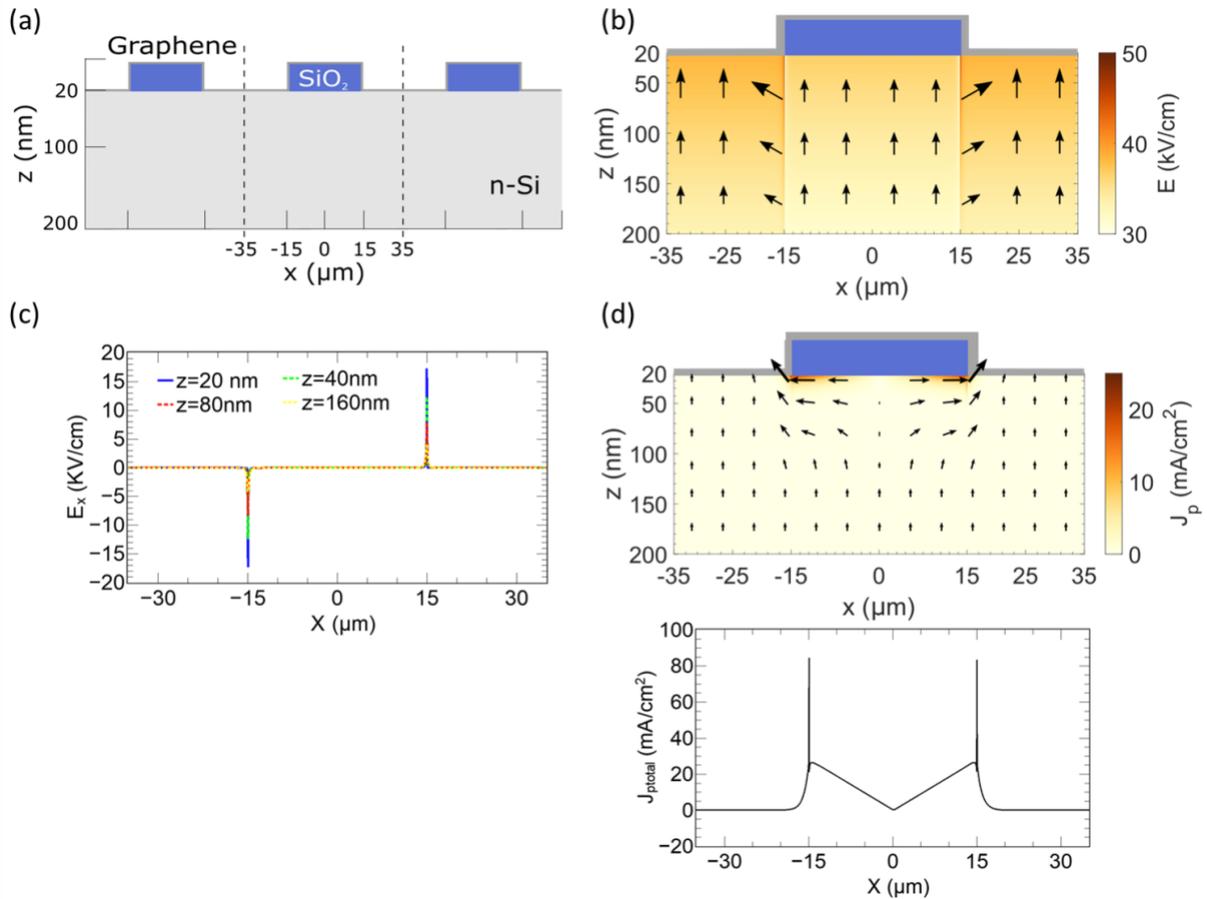

*Figure 4: (a) Schematics of the multifinger structure representing a cross section of device D1. Vertical dashed lines enclose one finger from -35 μm to 35 μm, with horizontal axis X = 0 μm located at the center of the finger. SiO$_2$ thickness is 20 nm and z = 20 nm corresponds to the SiO$_2$/Si interface. (b) Total electric field in the simulated structure. (c) Lateral electric field (E$_x$) at four different z locations: 20 nm, 40 nm, 80 nm and 160 nm respectively. Positive and negative values indicate electric fields aiming to the right or left, respectively. (d) Hole current density under uniform illumination with light of λ = 800 nm and intensity of 25 μW/cm$^2$ Black arrows indicate the direction and magnitude of J$_p$. The lower half depicts the total current (J$_p$) at z = 20 nm along the simulated structure.*

Under these conditions, the total current flowing in the device was numerically calculated. Fig. 4d presents the hole current density distribution (J$_p$) in the simulated finger, showing that its highest contribution appears close to the GIS – G/Si junction. The holes photogenerated below the insulator are attracted to the surface by the electric field previously depicted, but as they encounter the barrier created by the insulator (tunneling is not allowed) they move laterally to the nearest G/Si Schottky contact. Black arrows represent the direction and magnitude of the hole current. Due to the symmetry of the structure, the lateral contribution of $J_p$ in the center is minimum compared with the rest of the finger. The lower portion of Fig. 4d depicts the total hole current ($J_p$) at z = 30 nm. $J_p$ includes a vertical ($J_{pz}$) and a lateral component ($J_{px}$) of the hole current and it is equal to $J_p = \sqrt{J_{px}^2 + J_{pz}^2}$. An increasing value of the current below the SiO$_2$ from the center to the edges can be seen, which is equal to the lateral hole current ($J_{px}$). The vertical hole current ($J_{pz}$), in contrast, is zero below the insulator and at its maximum at the G/Si region closest to the GIS junction (see Fig. S6b). This high value of the injected current has its origin in the photogenerated holes below the insulator and represents the current injected into the G/Si contact (measured current). The hole current value then decreases just a few microns outside the GIS region and relaxes towards a constant value at the distance of about Δx = 5 µm from the SiO$_2$ finger edges (lower part of Fig. 4d and Fig. S6b). This value corresponds to the photogenerated holes below the G/Si contact and Δx provides an estimation of the width of the G/Si Schottky contact necessary to extract the photocurrent generated below the SiO$_2$ finger.

Based on the experimental observations and simulation results, qualitative energy band diagrams of D1 are proposed in Fig. 5. The lateral cross-section of the device along the length (x) and depth (z) dimensions under reverse applied bias is presented in Fig. 5a. The energy band diagrams of the device at the G/Si and GIS regions under dark conditions are shown in Fig. 5b. When the reverse bias voltage is increased above V$_{th}$, the depletion width of n-Si under graphene widens, whereas an inversion layer is formed in the n-Si substrate under the G/SiO$_2$ region, which limits the extension of the depletion region below the insulator.[3] The inversion layer causes a difference in the band alignment between the GIS and G/Si junctions along the n-Si substrate in the x-direction (Fig. 5c). Consequently, a lateral electric field along the x direction appears in the transition region between both junctions, driving minority charge carriers (i.e. holes) from the GIS to the G/Si region. Under this condition, when the device is illuminated, the photogenerated holes in the G/Si Schottky junction are extracted directly through the graphene contact. However, in the GIS junction, photo generated holes need to traverse the distance from the point of generation to the G/Si Schottky junction, because the 20 nm thick oxide rules out the possibility of a significant tunnel current through the $SiO_2$. Fig. 5d shows a schematic diagram depicting generation, extraction and collection of photogenerated charges carriers in one of the GIS fingers, according to the proposed band diagrams based on simulation studies.

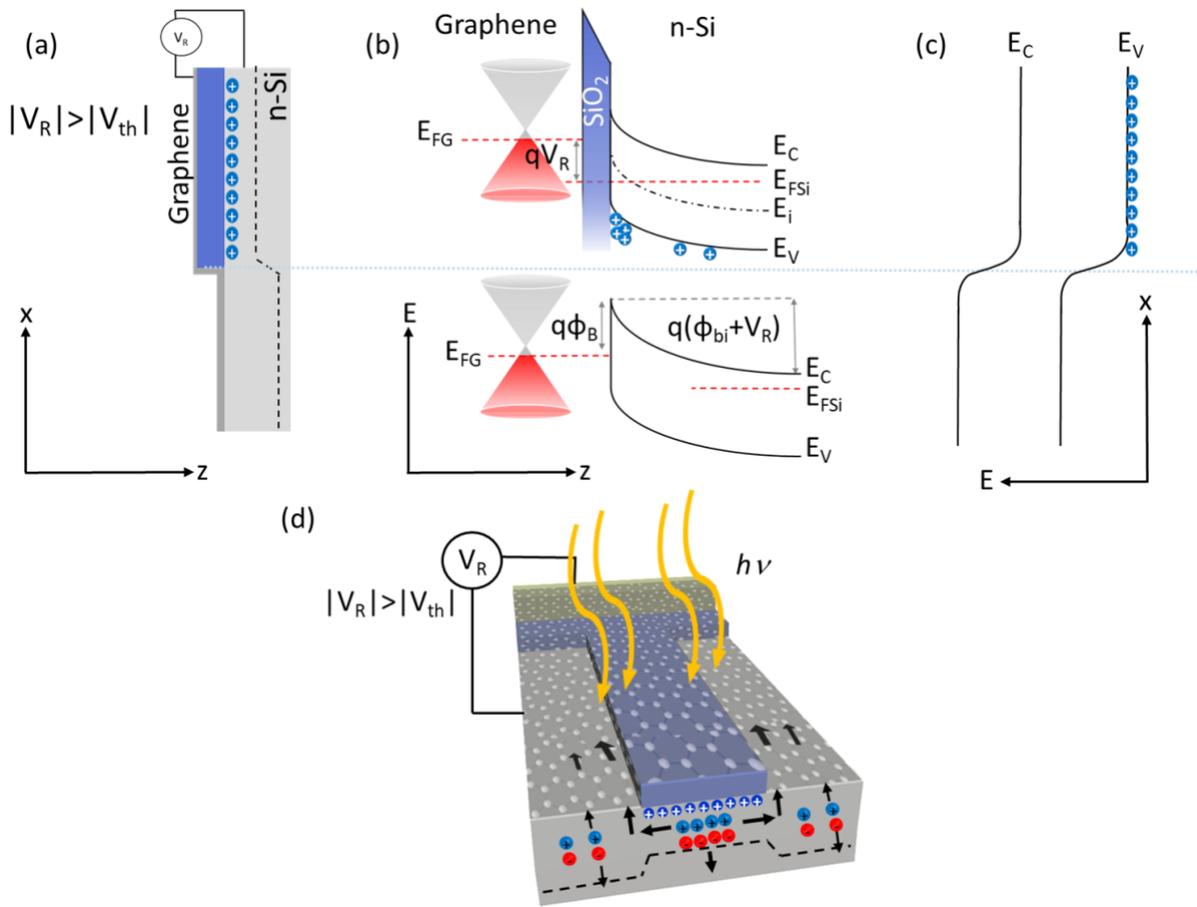

*Fig. 5: (a) Cross section of the device containing both G/Si and GIS regions at reverse bias above threshold voltage. Dashed line in n-Si substrate represents the limit/edge of the depletion region. (b) Schematic band diagram of the GIS (above) and G/Si (below) junctions along the z axis in dark conditions. (c) Lateral band diagram along the surface of n-Silicon (x-axis), just at the top interface with graphene and $SiO_2$ in reverse biased conditions. $E_C$, $E_V$, $E_{FSi}$, $E_i$, $E_{FG}$, $\Phi_{bi}$, $\Phi_B$ and $V_R$ indicate conduction band, valence band, Fermi level, intrinsic energy level of Si, graphene Fermi level, built-in potential, Schottky barrier height (SBH), and*

*reverse bias voltage of the diode, respectively. (d) Schematic of one of the fingers under illumination depicting the generation, extraction and collection of photogenerated charge carriers (larger size). Photogenerated electrons and holes in Si are shown in red and dark blue, respectively.*

The numerical study provides strong evidence that charge carriers photogenerated below the SiO$_2$ fingers first drift to the SiO$_2$/Si interface and then diffuse laterally up to the transition region with the G/Si junction where a lateral field accelerates them towards the Schottky contact. Figs. 5c and S4 show that the bands are bent laterally at the transition between junctions and become flat inside the GIS region. This suggests that the GIS regions should be patterned with dimensions such that the photogenerated holes can be extracted effectively and efficiently, thereby making the width of interdigitated GIS region a critical parameter for improved performance of G/Si photodiodes. In this respect, we fabricated devices with varying widths of the GIS regions up to 100 µm (see Fig. S1). We found that the most efficiently designed structure is device D1 with the highest SR and EQE values, in which the GIS regions have been patterned to a width of 30 µm. When the width of the GIS region is increased up to 100 µm, the responsivity drops to around 500 mA/W, but still exhibits a higher photoresponse when compared to the conventional structure, i.e. device D2. The decrease in responsivity for devices with fingers wider than D1 is explained by the fact that photogenerated carriers in the Si under G/SiO$_2$ are far away from G/Si junction and thus encounter a weaker or no lateral electric field. This increased the probability for carrier recombination before they can be extracted through the Schottky junction.

**Conclusions**

We demonstrated enhanced performance of G/Si photodiodes over a broad range of wavelengths. This was achieved through a simple design of interdigitated regions, where Schottky diodes are in parallel to regions where graphene is separated from the silicon by an insulating oxide. This smart contact design with graphene as a transparent electrode enables the efficient extraction of photogenerated charge carriers and resulted in a maximum EQE of 98% and responsivity of 635 mA/W. Simulations confirmed the formation of an inversion layer under the insulated regions, which serve to passivate defects and limit recombination of photogenerated carriers. The extraction of minority carriers, once the inversion layer is created, is efficiently enhanced by the stronger lateral electric fields derived by defining the G/Si contact geometry. Given the simple planar geometry of the fabricated devices and the availability of large-area CVD-grown graphene, the proposed fabrication process can easily be integrated into conventional semiconductor technology. The proposed interdigitated contact strategy can be further extended to other 2D-3D and 2D-2D heterostructures where graphene is used as one of the electrodes and semiconductors other than Si are used.

**Methods**

**Device Fabrication:**

A phosphorus-doped Si wafer with a nominal doping concentration of $2\times10^{15}$ cm$^{-3}$ was used as substrate. A thermal oxidation process yielded 20 nm of high-quality SiO$_2$ on the silicon surface. For chip fabrication, the wafer was diced into 20×20 mm$^2$ samples. The oxide was etched with hydrofluoric acid (HF) after a first standard UV-photolithography step in order to expose the n-Si substrate. The contact metal electrodes were defined by a second photolithography step followed by sputtering of 15 nm chromium (Cr) and 85 nm nickel (Ni) and liftoff process. The metal electrodes were deposited immediately after the native oxide removal, ensuring the formation of ohmic contacts. Large-area graphene was grown on a copper foil in a NanoCVD (Moorfield, UK) rapid thermal processing tool. To transfer graphene films onto pre-patterned substrates, ~ 7 mm$^2$ pieces of graphene-coated Cu foil were spin-coated with Poly methyl methacrylate (PMMA) and baked on a hot plate at 85°C for 5 minutes. A wet etch process was used to remove the polymer-supported graphene films from the copper surface. Iron (III) chloride (FeCl3) was used as an etchant to remove the copper. The native SiO$_2$ on the n-Si substrates was removed by HF prior to the graphene transfer to achieve an electrical contact between graphene and n-Si substrate. Afterwards, the chip was baked at 180°C for 35 min and then thoroughly immersed into acetone for one hour, followed by cleaning with isopropanol and drying. Then, a last photolithography step was performed, followed by oxygen plasma etching of graphene to define the graphene areas. For packaging and wirebonding, the patterned chip was diced into the 6×6 mm$^2$ samples. Before the dicing step, the chip was spin-coated with AZ 5214 E photoresist to protect the photodiodes during the dicing process. This layer was removed after dicing by immersion in acetone. Finally, the chips were glued into a chip package and the devices were wirebonded.

**Electrical Characterization:**

The chip package was inserted into a socket located inside a home-built chamber connected to a Keithley semiconductor analyzer (SCS4200) for electrical measurements under ambient condition. The voltage for all devices was swept from 0 to +2 V for forward ($V_F$) and from 0 to -2V for reverse ($V_R$) biasing. A white light source (LED lamp) with a dimmer to control the light intensity was used to confirm that the diodes are generally sensitive to light. The LED light spectrum can be found in supplementary Fig. S6. The intensity of the light source was measured by a CA 2 laboratory thermopile.

**Optical Characterization:**

The spectral response (SR) of the photodetectors was measured by comparing it to a calibrated Si reference diode using a lock-in technique. A tungsten-halogen lamp with a wavelength ranging from 300 nm to 2000 nm was used as the light source. Specific wavelengths were selected by a monochromator. The intensity of the light beam was modulated by a chopper with a frequency of 17 Hz. The current was measured by pre-amplifiers and lock-in amplifiers with 300 ms integration time and 0.4 Hz bandwidth at 17 Hz chopper frequency for detection of low currents down to 10 pA.

**Scanning photocurrent measurements:**

Scanning photocurrent measurements were performed using a Witec Alpha300 R confocal microscope equipped with a piezoelectric scanning stage. The microscope was coupled to a 532 nm wavelength laser to generate spatially resolved photocurrent, which is converted into a voltage signal using a current preamplifier and is recorded by a lock-in amplifier.

**Numerical simulation:**

Simulations of the structures under investigation have been carried out by self-consistently solving 2D electrostatic and transport equations. In particular we solve:

- Poisson's equation:

$$\nabla \cdot (\varepsilon \nabla \psi) = -q(p - n + N_D^+ - N_A^-)$$

where $\varepsilon$ is the dielectric constant; $\psi$ is the electrostatic potential, $q$ is the electron charge, $p$ ($n$) corresponds to the hole (electron) density, and $N_D^+$ ($N_A^-$) to the ionized donor (acceptor) doping density.

- Carrier continuity equation:

$$\frac{\partial n}{\partial t} = \frac{1}{q}\nabla \vec{J_n} + G_n - R_n; \qquad \frac{\partial p}{\partial t} = -\frac{1}{q}\nabla \vec{J_p} + G_p - R_p$$

where $G_n$ ($G_p$) and $R_n$ ($R_p$) are the electron (hole) generation and recombination terms, respectively. $J_n$ and $J_p$ are the electron and hole Drift-Diffusion equations expressed as:

$$\vec{J_n} = qn\mu_n \nabla\psi + qD_n \nabla n; \qquad \vec{J_p} = qp\mu_p \nabla\psi - qD_p \nabla p$$

where $\mu_n$ ($\mu_p$) is the electron (hole) mobility and $D_n$ ($D_p$) is the electron (hole) diffusion coefficient. Stationary conditions have been considered for this work.

Carrier recombination is expressed as:

$$R_{SRH} = \frac{np - n_i^2}{\tau_n(n + n_i^2) + \tau_p(p + p_i^2)}$$

with $\tau_n$ ($\tau_p$) the electron (hole) lifetimes, and $n_i$ ($p_i$) the intrinsic electron (hole) concentration. $\tau_n = \tau_p$ was assumed in this work.

Generation of electron-hole pairs is produced by light absorption that follows the Beer-Lambert's law:

$$G_n = G_p = \frac{P_{opt}}{E_{ph}} \alpha e^{-\alpha x}$$

where $P_{opt}$ is the applied light power at the surface in W/cm², $E_{ph} = \frac{hc}{\lambda}$ is the photon energy, and $\alpha$ is the absorption coefficient.

The current injection at the Schottky contacts has been modeled as:

$$J_n = -qS_n(n - n_0); \quad J_p = qS_p(p - p_0)$$

where $S_n$ ($S_p$) is the electron (hole) surface recombination velocity, and $n_0$ ($p_0$) is the electron (hole) density at the interface at equilibrium. $S_n$ ($S_p$) describes the electron (hole) transfer rate between the Si-substrate and the graphene contact. The value of this parameter is affected by the quality of both materials, the presence of impurities or an insulator barrier at the interface produced by the growing of a native SiO₂ [35].

**Acknowledgements**

Funding from the German Research Foundation (DFG GRK 1564) and the European Commission through the project Graphene Flagship (785219) is gratefully acknowledged. This work has been partially supported by the Spanish Ministry of Education, Culture and Sport (MECD), and the University of Granada through the project TEC2017-89955-P, the pre-doctoral grant FPU014/02579.

# References

(1) Nair, R. R.; Blake, P.; Grigorenko, A. N.; Novoselov, K. S.; Booth, T. J.; Stauber, T.; Peres, N. M. R.; Geim, A. K. Fine Structure Constant Defines Visual Transparency of Graphene. *Science* **2008**, *320* (5881), 1308.

(2) Di Bartolomeo, A. Graphene Schottky Diodes: An Experimental Review of the Rectifying Graphene/Semiconductor Heterojunction. *Phys. Rep.* **2016**, *606* (Supplement C), 1–58.

(3) Riazimehr, S.; Kataria, S.; Bornemann, R.; Haring Bolívar, P.; Ruiz, F. J. G.; Engström, O.; Godoy, A.; Lemme, M. C. High Photocurrent in Gated Graphene–Silicon Hybrid Photodiodes. *ACS Photonics* **2017**, *4* (6), 1506–1514.

(4) Riazimehr, S.; Bablich, A.; Schneider, D.; Kataria, S.; Passi, V.; Yim, C.; Duesberg, G. S.; Lemme, M. C. Spectral Sensitivity of Graphene/Silicon Heterojunction Photodetectors. *Solid-State Electron.* **2016**, *115* (Part B), 207–212.

(5) Pospischil, A.; Humer, M.; Furchi, M. M.; Bachmann, D.; Guider, R.; Fromherz, T.; Mueller, T. CMOS-Compatible Graphene Photodetector Covering All Optical Communication Bands. *Nat. Photonics* **2013**, *7* (11), 892.

(6) Pasternak, I.; Wesolowski, M.; Jozwik, I.; Lukosius, M.; Lupina, G.; Dabrowski, P.; Baranowski, J. M.; Strupinski, W. Graphene Growth on Ge(100)/Si(100) Substrates by CVD Method. *Sci. Rep.* **2016**, *6*, 21773–21773.

(7) Dragoman, D.; Dragoman, M.; Plana, R. Graphene-Based Ultrafast Diode. *J. Appl. Phys.* **2010**, *108* (8), 084316.

(8) Duesberg, G. S.; Kim, H. Y.; Lee, K.; McEvoy, N.; Winters, S.; Yim, C. Investigation of Carbon-Silicon Schottky Diodes and Their Use as Chemical Sensors. In *2013 Proceedings of the European Solid-State Device Research Conference (ESSDERC)*; 2013; pp 85–90.

(9) Kim, H.-Y.; Lee, K.; McEvoy, N.; Yim, C.; Duesberg, G. S. Chemically Modulated Graphene Diodes. *Nano Lett.* **2013**, *13* (5), 2182–2188.

(10) Wang, T.; Huang, D.; Yang, Z.; Xu, S.; He, G.; Li, X.; Hu, N.; Yin, G.; He, D.; Zhang, L. A Review on Graphene-Based Gas/Vapor Sensors with Unique Properties and Potential Applications. *Nano-Micro Lett.* **2016**, *8* (2), 95–119.

(11) Li, X.; Lv, Z.; Zhu, H. Solar Cells: Carbon/Silicon Heterojunction Solar Cells: State of the Art and Prospects (Adv. Mater. 42/2015). *Adv. Mater.* **2015**, *27* (42), 6767–6767.

(12) Song, Y.; Li, X.; Mackin, C.; Zhang, X.; Fang, W.; Palacios, T.; Zhu, H.; Kong, J. Role of Interfacial Oxide in High-Efficiency Graphene–Silicon Schottky Barrier Solar Cells. *Nano Lett.* **2015**, *15* (3), 2104–2110.

(13) An, X.; Liu, F.; Kar, S. Optimizing Performance Parameters of Graphene–Silicon and Thin Transparent Graphite–Silicon Heterojunction Solar Cells. *Carbon* **2013**, *57* (Supplement C), 329–337.

(14) Miao, X.; Tongay, S.; Petterson, M. K.; Berke, K.; Rinzler, A. G.; Appleton, B. R.; Hebard, A. F. High Efficiency Graphene Solar Cells by Chemical Doping. *Nano Lett.* **2012**, *12* (6), 2745–2750.

(15) Amirmazlaghani, M.; Raissi, F.; Habibpour, O.; Vukusic, J.; Stake, J. Graphene-Si Schottky IR Detector. *IEEE J. Quantum Electron.* **2013**, *49* (7), 589–594.

(16) Li, X.; Zhu, M.; Du, M.; Lv, Z.; Zhang, L.; Li, Y.; Yang, Y.; Yang, T.; Li, X.; Wang, K.; et al. High Detectivity Graphene-Silicon Heterojunction Photodetector. *Small Weinh. Bergstr. Ger.* **2016**, *12* (5), 595–601.

(17) Chen, Z.; Cheng, Z.; Wang, J.; Wan, X.; Shu, C.; Tsang, H. K.; Ho, H. P.; Xu, J.-B. High Responsivity, Broadband, and Fast Graphene/Silicon Photodetector in Photoconductor Mode. *Adv. Opt. Mater.* **2015**, *3* (9), 1207–1214.

(18) Wang, X.; Cheng, Z.; Xu, K.; Tsang, H. K.; Xu, J.-B. High-Responsivity Graphene/Silicon-Heterostructure Waveguide Photodetectors. *Nat. Photonics* **2013**, *7* (11), 888.

(19) Lv, P.; Zhang, X.; Deng, W.; Jie, J. High-Sensitivity and Fast-Response Graphene/Crystalline Silicon Schottky Junction-Based Near-IR Photodetectors. *IEEE ELECTRON DEVICE Lett.* **2013**, *34* (10), 1337–1339.

(20) Srisonphan, S. Hybrid Graphene–Si-Based Nanoscale Vacuum Field Effect Phototransistors. *ACS Photonics* **2016**, *3* (10), 1799–1808.

(21) Bartolomeo, A. D.; Luongo, G.; Giubileo, F.; Funicello, N.; Niu, G.; Thomas Schroeder; Lisker, M.; Lupina, G. Hybrid Graphene/Silicon Schottky Photodiode with Intrinsic Gating Effect. *2D Mater.* **2017**, *4* (2), 025075.

(22) Tao, L.; Chen, Z.; Li, X.; Yan, K.; Xu, J.-B. Hybrid Graphene Tunneling Photoconductor with Interface Engineering towards Fast Photoresponse and High Responsivity. *Npj 2D Mater. Appl.* **2017**, *1* (1).

(23) Luongo, G.; Giubileo, F.; Genovese, L.; Iemmo, L.; Martucciello, N.; Di Bartolomeo, A. I-V and C-V Characterization of a High-Responsivity Graphene/Silicon Photodiode with Embedded MOS Capacitor. *Nanomaterials* **2017**, *7* (7).

(24) An, Y.; Behnam, A.; Pop, E.; Ural, A. Metal-Semiconductor-Metal Photodetectors Based on Graphene/p-Type Silicon Schottky Junctions. *Appl. Phys. Lett.* **2013**, *102* (1), 013110.

(25) Goykhman, I.; Sassi, U.; Desiatov, B.; Mazurski, N.; Milana, S.; de Fazio, D.; Eiden, A.; Khurgin, J.; Shappir, J.; Levy, U.; et al. On-Chip Integrated, Silicon–Graphene Plasmonic Schottky Photodetector with High Responsivity and Avalanche Photogain. *Nano Lett.* **2016**, *16* (5), 3005–3013.

(26) Liu, F.; Kar, S. Quantum Carrier Reinvestment-Induced Ultrahigh and Broadband Photocurrent Responses in Graphene–Silicon Junctions. *ACS Nano* **2014**, *8* (10), 10270–10279.

(27) An, X.; Liu, F.; Jung, Y. J.; Kar, S. Tunable Graphene–Silicon Heterojunctions for Ultrasensitive Photodetection. *Nano Lett.* **2013**, *13* (3), 909–916.

(28) Bartolomeo, A. D.; Giubileo, F.; Luongo, G.; Iemmo, L.; Martucciello, N.; Niu, G.; Mirko Fraschke; Skibitzki, O.; Schroeder, T.; Lupina, G. Tunable Schottky Barrier and High Responsivity in Graphene/Si-Nanotip Optoelectronic Device. *2D Mater.* **2017**, *4* (1), 015024.

(29) Wan, X.; Xu, Y.; Guo, H.; Shehzad, K.; Ali, A.; Liu, Y.; Yang, J.; Dai, D.; Lin, C.-T.; Liu, L.; et al. A Self-Powered High-Performance Graphene/Silicon Ultraviolet Photodetector with Ultra-Shallow Junction: Breaking the Limit of Silicon? *Npj 2D Mater. Appl.* **2017**, *1* (1), 4.

(30) Kuruvila, A.; Kidambi, P. R.; Kling, J.; Wagner, J. B.; Robertson, J.; Hofmann, S.; Meyer, J. Organic Light Emitting Diodes with Environmentally and Thermally Stable Doped Graphene Electrodes. *J. Mater. Chem. C* **2014**, *2* (34), 6940–6945.

(31) Selvi, H.; Unsuree, N.; Whittaker, E.; Halsall, M. P.; Hill, E. W.; Thomas, A.; Parkinson, P.; Echtermeyer, T. J. Towards Substrate Engineering of Graphene–Silicon Schottky Diode Photodetectors. *Nanoscale* **2018**, *10* (7), 3399–3409.

(32) Gao, Y.; Cansizoglu, H.; Polat, K. G.; Ghandiparsi, S.; Kaya, A.; Mamtaz, H. H.; Mayet, A. S.; Wang, Y.; Zhang, X.; Yamada, T.; et al. Photon-Trapping Microstructures Enable High-Speed High-Efficiency Silicon Photodiodes. *Nat. Photonics* **2017**, *11* (5), 301–308.

(33) Smith, A. D.; Wagner, S.; Kataria, S.; Malm, B. G.; Lemme, M. C.; Östling, M. Wafer-Scale Statistical Analysis of Graphene FETs, Part I: Wafer-Scale Fabrication and Yield Analysis. *IEEE Trans. Electron Devices* **2017**, *64* (9), 3919–3926.

(34) Simon M. Sze; Kwok K. Ng. *Physics of Semiconductor Devices*, 3rd ed.; Wiley.

(35) Cheung, S. K.; Cheung, N. W. Extraction of Schottky Diode Parameters from Forward Current-voltage Characteristics. *Appl. Phys. Lett.* **1986**, *49* (2), 85–87.

(36) Chen, C.-C.; Aykol, M.; Chang, C.-C.; Levi, A. F. J.; Cronin, S. B. Graphene-Silicon Schottky Diodes. *Nano Lett.* **2011**, *11* (5), 1863–1867.

(37) Tongay, S.; Lemaitre, M.; Miao, X.; Gila, B.; Appleton, B. R.; Hebard, A. F. Rectification at Graphene-Semiconductor Interfaces: Zero-Gap Semiconductor-Based Diodes. *Phys. Rev. X* **2012**, *2* (1), 011002.

(38) Kim, S. M.; Hsu, A.; Lee, Y.-H.; Dresselhaus, M.; Palacios, T.; Kim, K. K.; Kong, J. The Effect of Copper Pre-Cleaning on Graphene Synthesis. *Nanotechnology* **2013**, *24* (36), 365602.

(39) Soylu, M.; Yakuphanoglu, F. Analysis of Barrier Height Inhomogeneity in Au/n-GaAs Schottky Barrier Diodes by Tung Model. *J. Alloys Compd.* **2010**, *506* (1), 418–422.

(40) Godfrey, R. B.; Green, M. A. A 15% Efficient Silicon MIS Solar Cell. *Appl. Phys. Lett.* **1978**, *33* (7), 637–639.

(41) Green, M. A.; Blakers, A. W. Advantages of Metal-Insulator-Semiconductor Structures for Silicon Solar Cells. *Sol. Cells* **1983**, *8* (1), 3–16.

(42) Juntunen, M. A.; Heinonen, J.; Vähänissi, V.; Repo, P.; Valluru, D.; Savin, H. Near-Unity Quantum Efficiency of Broadband Black Silicon Photodiodes with an Induced Junction. *Nat. Photonics* **2016**, *10* (12), 777–781.